\documentclass[fleqn,usenatbib,useAMS]{mnras}

\usepackage{graphicx}	
\usepackage{amsmath}	
\usepackage{amssymb}	
\usepackage{multicol}        
\usepackage{bm}		
\usepackage{pdflscape}	
\usepackage[T1]{fontenc}
\usepackage{ae,aecompl}
\usepackage{newtxtext,newtxmath}

\title[W vs Se emission lines in kilonova]{Tungsten vs Selenium as a potential source of kilonova nebular emission observed by Spitzer}

\author[K. Hotokezaka et al.]{
	Kenta Hotokezaka,$^{1,2}$\thanks{E-mail: kentah@g.ecc.u-tokyo.ac.jp}
	Masaomi Tanaka,$^{3,4}$ Daiji Kato,$^{5,6}$ 
	Gediminas Gaigalas$^{7}$
	\\
	$^1$Research Center for the Early Universe, Graduate School of Science, The University of Tokyo, Bunkyo, Tokyo 113-0033, Japan\\
	$^2$Kavli IPMU (WPI), UTIAS, The University of Tokyo, Kashiwa, Chiba 277-8583, Japan\\
	$^3$Astronomical Institute, Tohoku University, Sendai 980-8578, Japan
\\	
	$^4$Division for the Establishment of Frontier Sciences, Organization for Advanced Studies, Tohoku University, Sendai 980-8577, Japan
\\	
	$^5$National Institute for Fusion Science, 322-6 Oroshi-cho, Toki 509-5292, Japan\\
	$^6$Department of Advanced Energy Engineering Science, Kyushu University, Kasuga, Fukuoka 816-8580, Japan\\
	$^7$Institute of Theoretical Physics and Astronomy, Vilnius University, Saul\.{e}tekio Ave. 3, Vilnius, Lithuania}

\date{Accepted XXX. Received YYY; in original form ZZZ}

\pubyear{2022}

\begin{document}
\label{firstpage}
\pagerange{\pageref{firstpage}--\pageref{lastpage}}
\maketitle

\begin{abstract}
Infrared emission lines arising from transitions between fine structure levels of heavy elements are expected to produce  kilonova nebular emission. For the kilonova in GW170817, strong emission at $4.5\,{\rm \mu m}$ at late times was detected by the Spitzer Space Telescope   but no source was detected at $3.6\,{\rm \mu m}$. This peculiar spectrum indicates that there exist strong  line emitters around  $4.5\,{\rm \mu m}$ and the absence of strong lines around $3.6\,{\rm \mu m}$. To model the  spectrum  we prepare a line list based on the selection rules in LS coupling from the experimentally calibrated energy levels in the NIST database.
This method enables to generate the  synthetic spectra  with accurate line wavelengths. 
We find that the spectrum is  sensitive to the abundance pattern whether or not the first r-process peak elements are included. In both cases, the synthetic spectra can match the observed data, leading to two possible interpretations. If the first  peak elements are abundant a Se\,III line dominates the flux. If otherwise, W\,III with Os\,III, Rh\,III, and Ce\,IV can be the main sources. Observing nebular spectra for the future kilonovae in a wider  wavelength range can provide more conclusive elemental identification.
\end{abstract}


\section{Introduction}
The origin of r-process elements is a long-standing problem in astrophysics \citep{Burbidge1957RvMPB,Cameron1957PASP}.
Neutron star mergers have been considered as promising sites of r-process nucleosynthesis \citep{Lattimer1974ApJ}.
 A neutron star merger, GW170817, was accompanied by 
an uv-optical-infrared counterpart referred to as `kilonova' or `macronova' \citep{Abbott2017ApJ}.  The light curve and spectrum indicate that a large amount of r-process elements is produced in this event \citep[see][for reviews]{Metzger2017LRR,Nakar2020PhR,Margutti2021ARA&A}. 
The amount of the ejecta together with the event rate suggests that neutron star mergers can provide all the r-process elements in the Galaxy (e.g. \citealt{Hotokezaka2018IJMPD,Rosswog2018A&A}).

Currently, strontium (Sr, atomic number $Z=38$) seems only the element that is identified 
from a structure like the P Cygni profile in the early kilonova spectra (\citealt{Watson2019Natur} and see also \citealt{Domoto2021ApJ,Gillanders2022} but also \citealt{Perego2022ApJ} for an alternative explanation by He).
Apart from Sr, the spectral evolution implies the existence of 
lanthanides \citep{Tanaka2017PASJ,Kasen2017,kawaguchi2018,Wollaeger2021ApJ} because 
the opacity is sensitive to their abundances \citep{barnes2013ApJ,tanaka2013,Fontes2020MNRAS}.

Kilonova nebular emission at the later times imprints information on  atomic species synthesized in neutron star mergers. 
A fraction of nebular emission is expected to be 
radiated in infrared wavelengths through forbidden lines \citep{Hotokezaka2021MNRAS}.
The {\it Spitzer} Space Telescope 
observed the nebular phase of the kilonova in GW170817 at 43 and 74 days \citep{Kasliwal2022MNRAS}.
Strong emission was indeed detected at the $4.5\,{\rm \mu m}$ band \citep{Villar2018ApJ,Kasliwal2022MNRAS}, of which the amount of energy radiated in this band is about ten per cent of the total radioactive heating \citep[e.g.,][]{Hotokezaka2020ApJ}.
On the contrary, upper limits were obtained at the $3.6\,{\rm \mu m}$ band at both epochs, indicating that there exist strong emission lines around $4.5\,{\rm \mu m}$ and the absence of such lines in the $3.6\,{\rm \mu m}$ band. 
Although the physical conditions of kilonova nebulae are studied in the literature \citep{Hotokezaka2021MNRAS,Pognan2022MNRAS} any attempts to model this peculiar emission feature have not yet been made  because of the lack of atomic data with accurate line wavelengths.
In this {\it Letter}, we present the synthetic spectra of kilonova nebular emission in the wavelength range observed by  {\it Spitzer} including most of the relevant atomic species. 
In \S 2, we  estimate  the infrared emission  of fine structure lines of heavy elements. In \S 3, we produce a line list from the energy levels in the NIST database and present the synthetic spectra of kilonova nebular emission.  In \S 4, we conclude our findings.

\section{Fine structure emission lines}
The energy scale of fine structure splitting of heavy elements is typically $\sim 0.01$ -- $1$ eV. 
Because the optical depth of the kilonova ejecta to bound-bound transitions
is $\lesssim 0.1$ at $40\,{\rm days}$ at wavelengths longer than near infrared \citep[e.g.][]{Tanaka2020MNRAS,Pognan2022},  emission lines arising from  transitions between fine structure levels are expected to freely escape on this time scale. 
The strength of each line is determined by the level population and radiative transition rate.
The  rate of a magnetic dipole (M1) transition from an upper level $u$ to a lower level $l$  that satisfy the M1 selection rules in LS coupling\footnote{The selection rules in LS coupling are (1) $\Delta J=0,\,\pm 1$ (except $0\leftrightarrow0$), (2) no parity change, (3) no change in electron configuration, (4) $\Delta S=\Delta L=0$ and $\Delta J=\pm 1$.}  is given  by \citep{Pasternack1940ApJ,Shortly1940,Bahcall1968ApJ}
\begin{align}
    A_{ul}= 1.3\,{\rm s^{-1}}\,\left(\frac{\lambda_{ul}}{4\,{\rm \mu m}} \right)^{-3}f(J,L,S),\label{eq:M1}
\end{align}
where $\lambda_{ul}$ is the line wavelength and $f(J,L,S)$ is an algebraic factor depending on
the total angular momentum $J$, total orbital angular momentum $L$, and total spin angular momentum $S$ of the upper level in units of $\hbar$.  
LS coupling is known as a good approximation only 
for light elements. However, we find that this formula  is accurate to $10$ per cent even for 
a transition rate  of Ac III \citep[Z=89, ][]{NISTASD}.
The collisional deexcitation rate coefficient is given by
\begin{align}
    \langle \sigma v\rangle_{ul} = \frac{8.63\cdot 10^{-8}}{\sqrt{T_{e,4}}}\frac{\Omega_{ul}(T_e)}{g_u} \,{\rm cm^3 s^{-1}},
\end{align}
where $\Omega_{ul}(T_e)$ is the  collision strength, $g_u$ is the statistical weight of the upper level,
and $T_{e,4}$ is the electron temperature in units of $10^{4}\,{\rm K}$.
Comparing $A_{ul}$ and $\langle \sigma v\rangle_{ul}$ one finds a critical density of $\sim 10^{7}\,{\rm cm^{-3}}(g_u/\Omega_{ul})(T_e/3000\,{\rm K})^{1/2}(\lambda_{ul}/4{\rm \mu m)}^{-3}$. Because
the electron  density of the kilonova ejecta at 40 days is  $n_e \sim 10^5$ -- $10^6\,{\rm cm^{-3}}$ M1 radiative decay is typically faster than collisional deexcitation  suggesting that the ground levels are overpopulated compared to the expected values from the thermal distribution. Therefore   strong  lines in infrared wavelengths are expected to arise from  transitions between fine-structure levels in the ground terms. 

\begin{figure}
\begin{center}
\includegraphics[width=40mm]{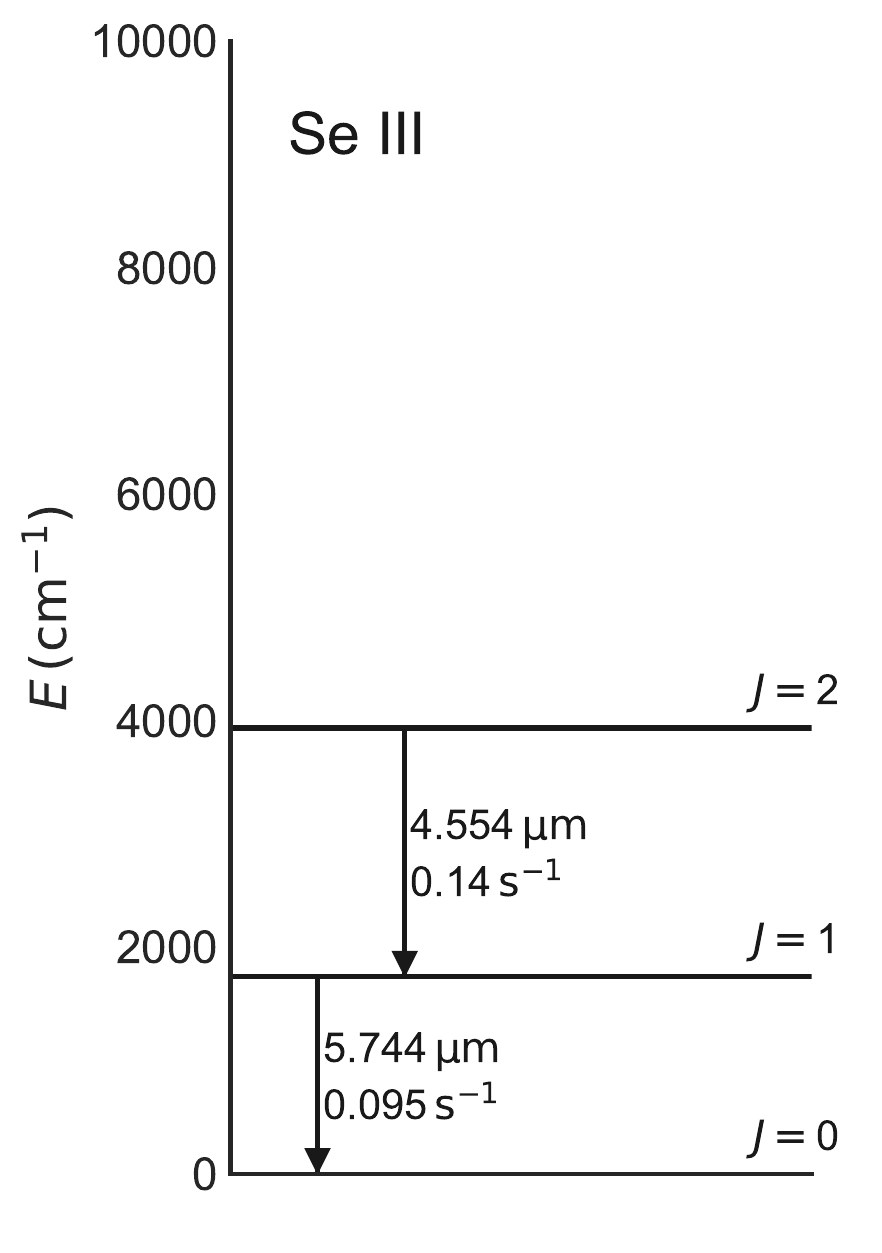}
\includegraphics[width=40mm]{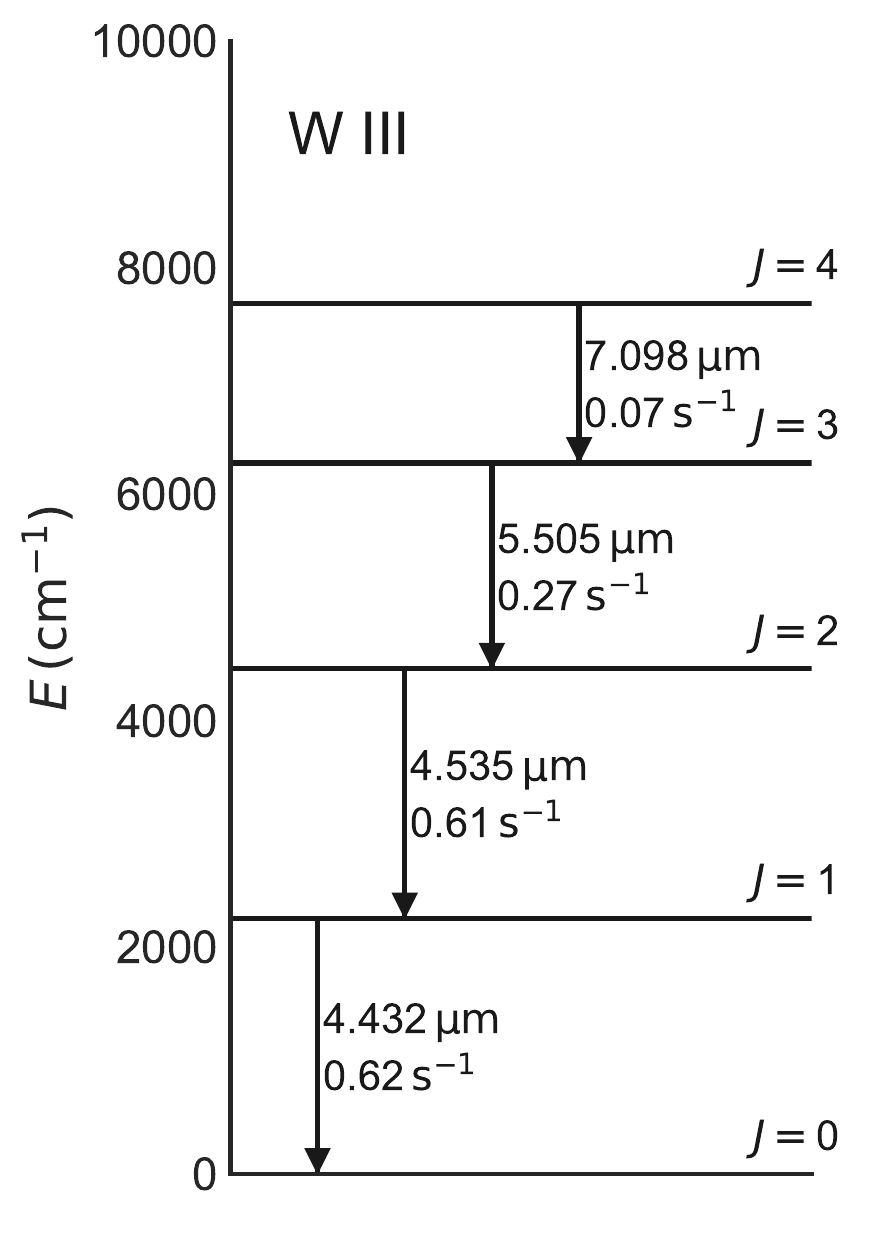}
\caption{Energy-level diagram of the  fine structure of the ground terms of Se III
4$p^2\, {\rm ^3P}_J$ ($Z=34$) and W III $5d^4~{\rm ^5D}_J$ ($Z=74$).
 Arrows depict M1 transitions with radiative transition rates and transition wavelengths.
 }
\label{fig:g1}
\end{center}
\end{figure}

\begin{figure}
\begin{center}
\includegraphics[width=80mm]{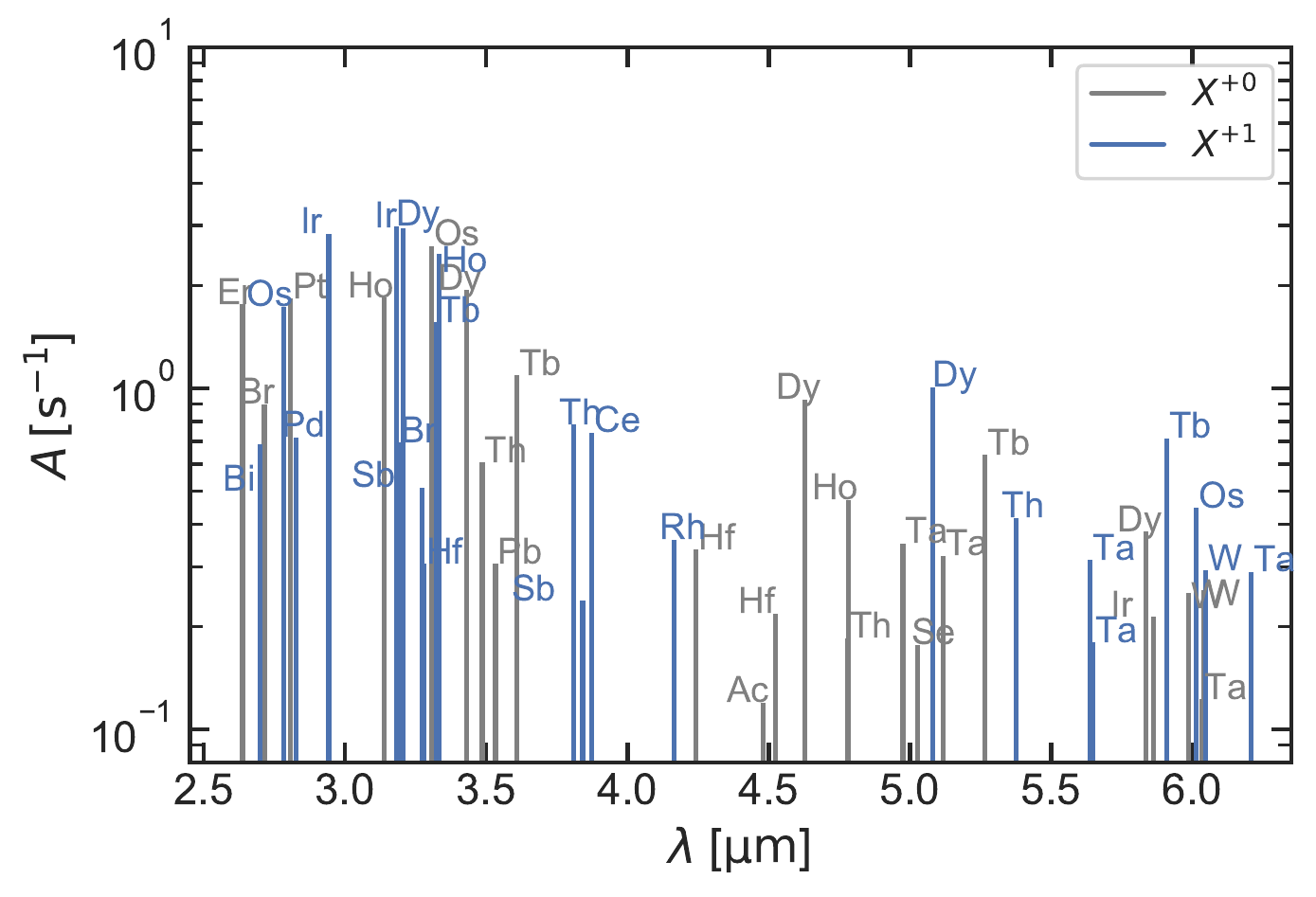}
\includegraphics[width=80mm]{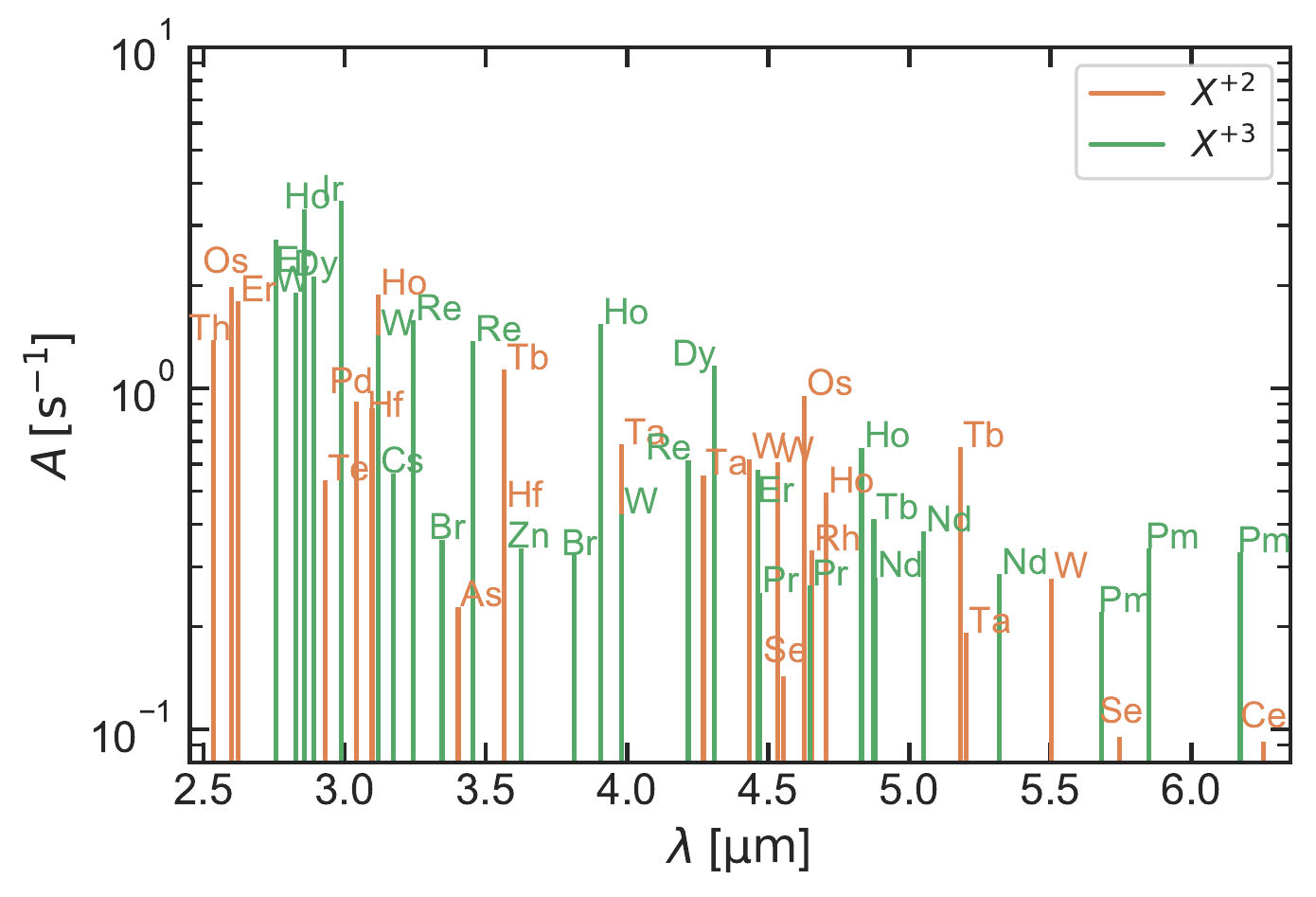}
\caption{M1 transitions between fine-structure levels of the ground terms
in the range of $2.5$ - $6.3\,{\rm \mu m}$. Ions, $X^{+0}$--$X^{+3}$, for  atomic numbers 
$Z\geq 30$ are shown. 
}
\label{fig:A}
\end{center}
\end{figure}


Doubly ionized selenium (Se III, $Z=34$) and tungsten (W III, $Z=74$) are good examples of ions having fine structure transitions in their ground term in the {\it Spitzer} $4.5\,{\rm \mu m}$ band.
Figure \ref{fig:g1} shows the fine structure levels of the ground terms $^3{\rm P}_J$ for Se III and $^5{\rm D}_J$  for W III \citep{NISTASD} with  arrows depicting  M1 transitions. Also shown are the radiative transition rates derived from equation \ref{eq:M1} and their wavelengths. Since ions with the ground term of $^5{\rm D}_{J}$ have five fine structure levels they are expected to be possible infrared line emitters  \citep{Berrington1995}.
W III indeed has two emission lines at $4.432$ and $4.535\,{\rm \mu m}$ from the first and the second excited levels.

At the densities lower than the critical density, 
the emission rate is determined by collisional excitation
rather than the radiative transition rates.
For a fine structure line $u\rightarrow l$,
the emission rate per ion is approximated by the collisional excitation rate from the ground level to $u$:
\begin{align}
    \epsilon_{ul} \approx \frac{0.0863}{\sqrt{T_{e,4}}} \frac{\Omega_{u0}}{g_0}\frac{n_e}{10^6\,{\rm cm^{-3}}}e^{-E_{u0}/kT_e}\,{\rm s^{-1}}\label{eq:line},
\end{align}
where $E_{u0}$ is the excitation energy of level $u$ from the ground level.

For instance, the total luminosity in  [W III]\,$4.43\,{\rm \mu m}$ and [W III]\,$4.54\,{\rm \mu m}$ for $T_e\sim 3500\,{\rm K}$ is estimated by
\begin{align}
    L_{4.5{\rm \mu m}}({\rm W\,III})\sim  5\cdot 10^{37}\,{\rm erg/s}\,
    \frac{\Omega_{u0} n_e}{10^6\,{\rm cm^{-3}}}
    \frac{M_{\rm ej}}{0.05M_{\odot}} 
    \frac{X_W}{0.004},
\end{align}
where $M_{\rm ej}$ is the total ejecta mass and
$X_W$ is the mass fraction of W III  including all the isotopes existing at a given epoch. 
It is possible that this  luminosity accounts for a large fraction of the observed luminosity in the  $4.5\,{\rm \mu m}$-band at $43$ days of $\Delta \nu L_{\nu}\sim 2\cdot 10^{38}\,{\rm erg/s}$ \citep{Kasliwal2022MNRAS} and $7\cdot 10^{37}\,{\rm erg/s}$ \citep{Villar2018ApJ}. 

In this paper, we assume that dust grains are absent in the kilonova ejecta, and therefore, the  infrared emission is entirely produced by atomic emission lines. This assumption is supported by \cite{Takami2014ApJ}, where it is shown that the dust formation in the ejecta is unlikely to occur unless light elements such as carbon are abundant \cite[see also][in the context of GW170817]{Gall2017ApJ,Villar2018ApJ}.

\section{Synthetic nebular spectrum for Spitzer observation}
In order to refine the rough estimates  given in the previous section we collected the experimentally calibrated energy levels that are available in the NIST database \citep{NISTASD} for neutral to triply ionized ions of all the relevant atomic species  and produced a line list  by using the M1 selection rules in the single configuration approximation. In particular, here we restrict ourselves to the LS  coupling scheme because the NIST database uses LS terms to represent energy levels for most ions relevant to this work.
In addition to the NIST data, the energy levels of Hf III and Ta III are
taken from \cite{Malcheva2009MNRAS} and \cite{Azarov2003PhyS},
of which the level identification may not be as accurate as the NIST data.
 The transition rate of each line is assigned according to the formula \ref{eq:M1}. On one hand, our line list is not necessarily complete because it is limited by the availability of energy levels in the NIST database and  lines existing in intermediate coupling are not included. On the other hand, this method guarantees good accuracy of wavelengths. Therefore it enables to generate  synthetic spectra, which are useful for comparison to observed data.
Figure \ref{fig:A} shows the fine-structure lines  of the ground terms of $X^{+0}$ -- $X^{+3}$ with  $Z\geq 30$ in  $2.5$ -- $6.3\,{\rm \mu m}$.   

Other key ingredients in the modelings are  collision strengths, $\Omega_{ul}$, which are generally large $\mathcal{O}(1)$ for fine structure transitions. In this work we use an atomic structure code \texttt{HULLAC} \citep{Bar-Shalom2001JQSR} to compute the collision strengths for transitions between  levels of  ground terms, which are the most relevant ones. Otherwise, we assume $\Omega_{ul}=1$. 

\begin{table*}
\begin{center}
\caption{{\it Spitzer} AB magnitude of the kilonova GW170817 at $43$ days.} \label{table:AB}
\vspace{-0.5cm}
\begin{tabular}{ccccccc} \hline \hline
 Wavelength & \texttt{model\_1st}, $X^{+1,+2}$  & \texttt{model\_1st}, $X^{+2,+3}$  & \texttt{model\_2nd-3rd},$X^{+1,+2}$  & \texttt{model\_2nd-3rd},$X^{+2,+3}$  & Observed$^{(1)}$ & Observed$^{(2)}$\\ \hline
$3.6\,{\rm \mu m}$ & 24.6 & 24.5 & 22.4 & 23.8 & $>23.21$ & $>23.3$\\
$4.5\,{\rm \mu m}$ &22.7 & 22.8  &22.8 & 22.7 & $21.88\pm 0.04 (\pm 0.05)$ & $22.9\pm 0.3$\\
 \hline \hline
\end{tabular}
\item[(1)] \cite{Kasliwal2022MNRAS}. The $3\sigma$ upper limit at $3.6\,{\rm \mu m}$ is shown. The systematic error is shown in the parenthesis. 
\item[(2)] \cite{Villar2018ApJ}. The $3\sigma$ upper limit  at $3.6\,{\rm \mu m}$ is shown. The error of the $4.5\,{\rm \mu m}$ detection is dominated by systematic uncertainties.
\end{center}
\end{table*}

Since the r-process abundance pattern of the ejecta in GW170817 is under debate, here we consider two sets of the abundance patterns: (1) the solar r-process abundance pattern \citep{Goriely1999A&A} including the first r-process peak elements (atomic mass numbers $A\geq 69$; \texttt{model\_1st}) and (2)  that excluding the first  peak ($A\geq 88$; \texttt{model\_2nd-3rd}). In the former case,  $\sim 70\%$ of the mass is in the first  peak elements. We assume the ejecta properties of $M_{\rm ej}=0.05M_{\odot}$, $v_{\rm ej}=0.1c$, $\chi_e=1$, and $T_e=2000\,{\rm K}$ for \texttt{model\_1st}, where $v_{\rm ej}$ and $\chi_e$ are the mean ejecta velocity and
free electron fraction.  For  \texttt{model\_2nd-3rd}, we use $M_{\rm ej}=0.05M_{\odot}$, $v_{\rm ej}=0.07c$, $\chi_e=2$, and $T_e=3500\,{\rm K}$.
The two examples of the spectra with the different ionization stages are considered: (i) $X^{+1}\&\,X^{+2}$ and (ii) $X^{+2}\&\,X^{+3}$. For demonstration purposes,
 the fractions of the two ionization stages are assumed to be equal in both cases. 
For a given $n_e$ and $T_e$, the level population of each ion is solved with a similar method employed in \cite{Hotokezaka2021MNRAS}.

\begin{figure*}
\begin{center}
\includegraphics[width=70mm]{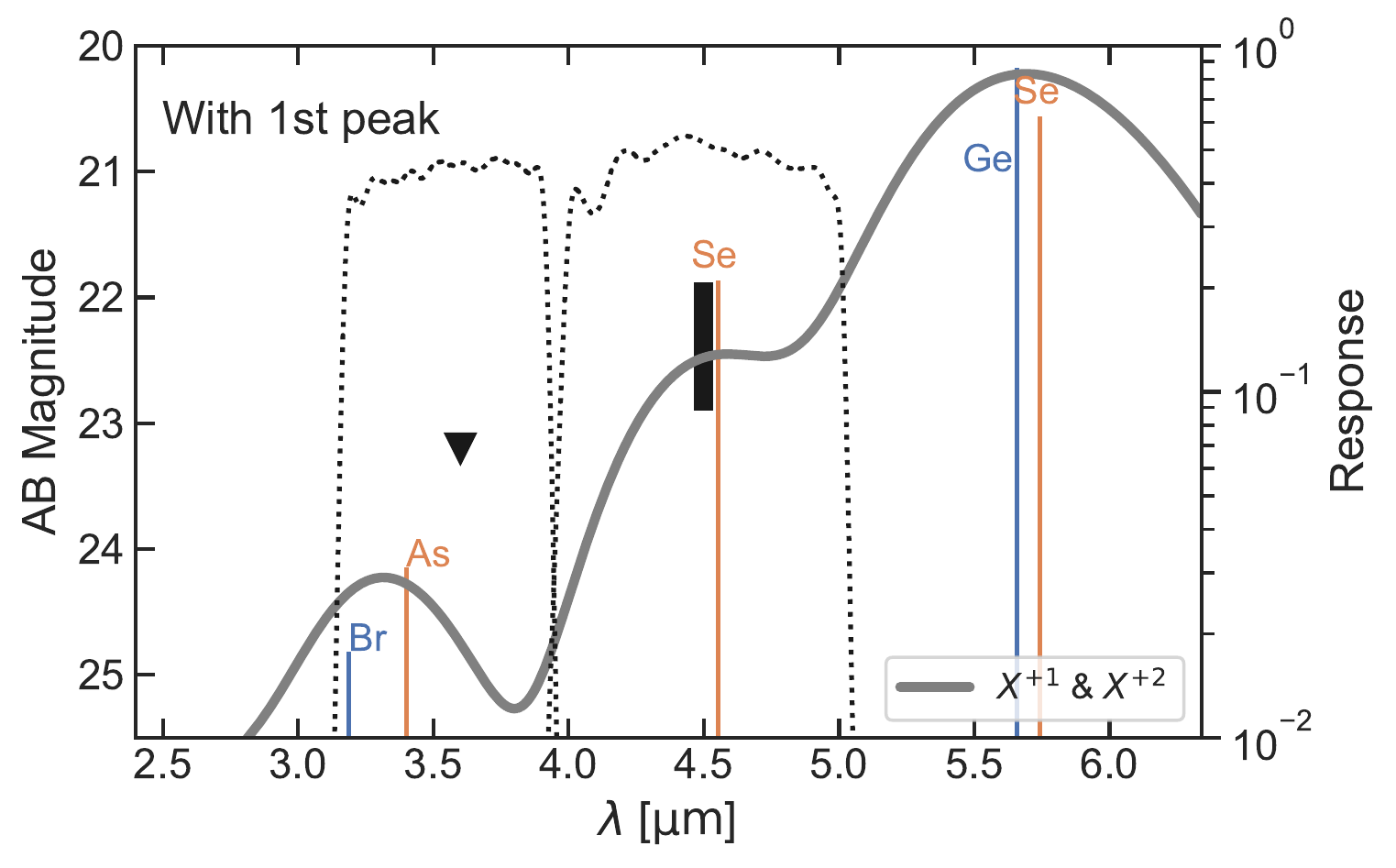}
\includegraphics[width=70mm]{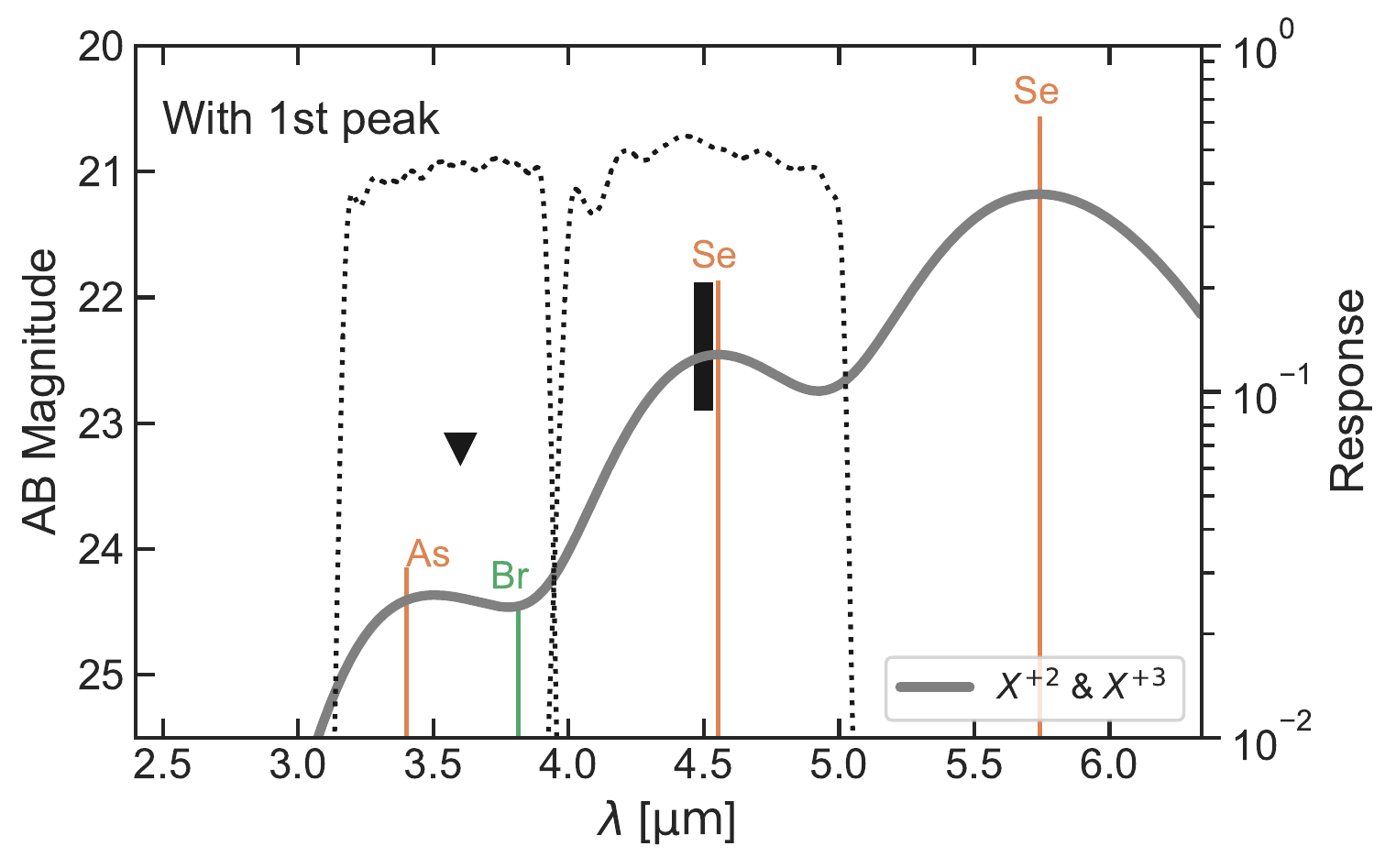}\\
\includegraphics[width=70mm]{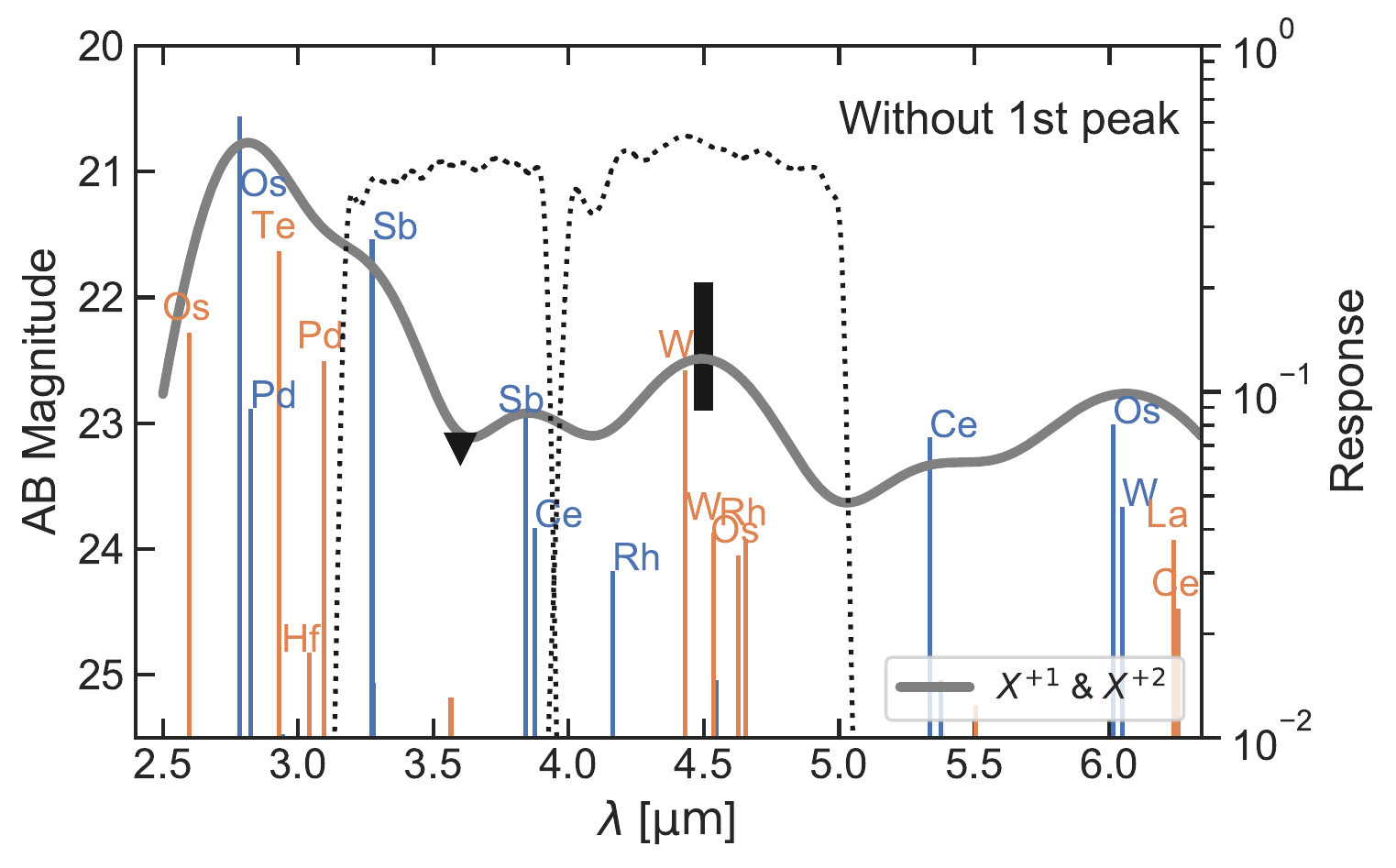}
\includegraphics[width=70mm]{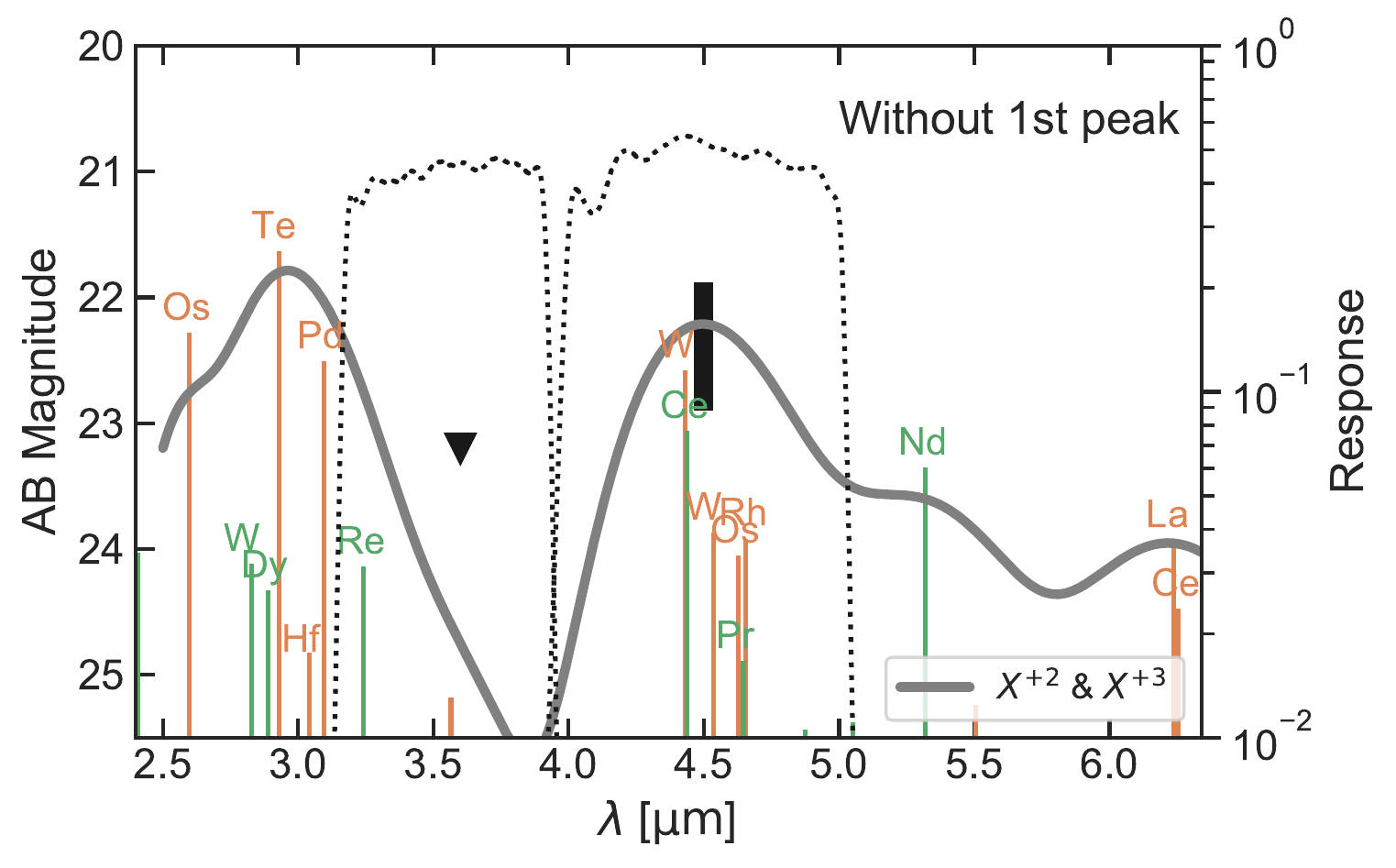}
\caption{Synthetic spectra for the kilonova in GW170817 at 43 days. 
The solar r-process abundance pattern with the first peak elements ($A\geq 69$, \texttt{model\_1st}) and that without the first peak elements ($A\geq 88$, \texttt{model\_2nd-3rd}) are used in the {\it upper} and {\it lower} panels.
The ionization fractions are assumed to be equal for $X^{+1}$ and $X^{+2}$ ({\it left})
and  for $X^{+2}$ and $X^{+3}$ ({\it right}). 
The total flux is obtained by adding the contributions of all lines, where the Doppler-shifted line profile is assumed to be a Gaussian in  frequency space with a broadening parameter $v_{\rm ej}=0.1c$ and $0.07c$ for \texttt{model\_1st} and \texttt{model\_2nd-3rd}, respectively. Blue, orange, and green vertical lines indicate the wavelength of each line in the rest frame for $X^{+1}$, $X^{+2}$, and $X^{+3}$, respectively. The height of the vertical lines corresponds to  the total flux in a line
divided by $v_{\rm ej}/\lambda_{ul}$.
Also depicted are the {\it Spitzer}-$4.5\,{\rm \mu m}$ detection, the $3.6\,{\rm \mu m}$ upper limit, and the spectral responses as a vertical bar,  a triangle, and dotted lines. (\url{https://irsa.ipac.caltech.edu/data/SPITZER/docs/irac/calibrationfiles/spectralresponse/}).
The bar indicates the range of the AB magnitudes estimated by the two different groups \citep{Kasliwal2022MNRAS,Villar2018ApJ}.
}
\label{fig:Sp}
\end{center}
\end{figure*}

Figure \ref{fig:Sp} shows the emission-line spectra at $43$ days.
 It is somewhat surprising that both \texttt{model\_1st}
 and \texttt{model\_2nd-3rd} can match the {\it Spitzer} detection and upper limit (see also table \ref{table:AB})
 except \texttt{model\_2nd-3rd} ($X^{+1}\&\,X^{+2}$)
 while their emission spectra  are clearly different outside the {\it Spitzer}-bands.
 In \texttt{model\_1st}, a single line, [Se III]\,${\rm 4.55\,{\rm \mu m}}$,  completely dominates the flux at the $4.5\,{\rm \mu m}$ band. This is simply because Se is  the most abundant atomic species. If this is the case,  an even stronger emission line [Se III]\,$5.74\,{\rm \mu m}$ must exist.
 Although As III, Br II, and Br IV, which are next to Se in the periodic table, have lines in the $3.6\,{\rm \mu m}$ band their fluxes are sufficiently small because of their lower abundances.
 In \texttt{model\_2nd-3rd},  W III, Ce IV, Rh III, and Os III can produce the observed $4.5\,{\rm \mu m}$ flux. However, Sb II overproduces emission in the $3.6\,{\rm \mu m}$ band compared to the observed limit.  
 This suggests that  most of Sb atoms are in highly ionized states, e.g., Sb III.

The observed luminosity in the $4.5\,{\rm \mu m}$ band at $43$ days is $\sim 10^{38}\,{\rm erg/s}$, which is a few per cent of or even comparable to the total luminosity expected from $\beta$-decay thermalization \citep[][]{Waxman2019ApJ,Kasen2019ApJ,Hotokezaka2020ApJ}. Such a high radiative efficiency in mid-infrared indicates that
the electron temperature in this epoch is low, $T_e \lesssim 5000\,{\rm K}$, otherwise
the thermal energy is predominantly radiated at the shorter wavelengths.
We estimate that the total luminosities in  M1 lines at $43\,{\rm days}$ are
 $\sim 5\cdot 10^{38}\,{\rm erg/s}$  for \texttt{model\_1st} and $\sim 2\cdot 10^{39}\,{\rm erg/s}$  for \texttt{model\_2nd-3rd}. In the case of \texttt{model\_1st} , the  luminosity  exceeds the total heating rate expected from the adopted abundances. This  indicates that the relative abundances of the first peak elements are not as large as the solar r-process abundances even if Se dominates the $4.5\,{\rm \mu m}$ flux.

It is also important to note that $\sim 50\%$ of W atoms at $43$ days are in radioactive isotopes $^{185}$W and $^{188}$W, of which half-lives are 75.1 and $69.8$ days, respectively. In \texttt{model\_2nd-3rd}, therefore, the relative strength of the emission around $4.5\,{\rm \mu m}$ is expected to decrease by a factor of $\sim 2$ after these isotopes have disappeared. The decay chain $^{188}$W$\rightarrow ^{188}$Re$\rightarrow^{188}$Os releases $\sim 900\,{\rm keV}$ in $\beta$ radiation, which means that W itself is an energy and ionizing source. Therefore,  the emission of W and Os may be enhanced if the light and heavy elements are spatially separated in the ejecta and 
$\beta$ electrons are locally confined by magnetic fields \citep{Barnes2016ApJ,Waxman2019ApJ}.
On the contrary, there are no radioactive elements for $69\leq A \leq 88$ with a  half-life of $10$  -- $100$ days, thereby the emission of Se may be strongly suppressed if such inhomogeneuous heating exists.

Our choices of the ionization stages are  motivated by  \cite{Hotokezaka2021MNRAS} and \cite{Pognan2022MNRAS}, where  ionization  in the kilonova nebular phase is studied.
 \cite{Hotokezaka2021MNRAS} showed that the dielectronic recombination rates of Nd ions are $\gtrsim 10^{-10}\,{\rm cm^3 s^{-1}}$, which are significantly larger than the radiative recombination rates. As a result, they found that Nd atoms are in Nd II and Nd III  around $40\,{\rm days}$. If the recombination rate is lower $\sim 10^{-11}\,{\rm cm^3 s^{-1}}$, the typical ionization stage is $X^{+2}$ or $X^{+3}$ \citep{Pognan2022MNRAS}.
The dielectronic recombination rates of Se and W ions are rather similar to Nd  at the nebular temperatures \citep{Sterling2011A&A,Preval2019JPhB}. Therefore we expect  that Se and W at $43$ days are predominantly in the singly and doubly ionized ions. If this is the case and \texttt{model\_2nd-3rd} is correct,   emission lines of [Os III]\,2.60\,${\rm \mu m}$, [Os II]\,2.78\,${\rm \mu m}$,
[Os II]\,6.01\,${\rm \mu m}$, and [W II]\,$6.05\,{\rm \mu m}$ may produce observable features.


We now turn to discuss the caveats in our analysis. Our line list is not necessarily complete because it contains only M1 transitions within the single configuration approximation, i.e., most lines in our list are those arising from transitions  between  fine structure levels  of a given LS term. This  is no longer a good approximation for heavy elements, where two types of mixing are important; (i) the mixing
between different final and intermediate terms within the same configuration and (ii) the mixing between different configurations. The former effect starts to appear for elements around the second r-process peak, $Z\gtrsim 50$. The latter is important for heavier elements, e.g., $Z\gtrsim 74$. 
Note, however, that fine structure transitions are usually less sensitive to these effects compared to other transitions such as optical lines. 
 Our list is also limited by the availability of  energy levels in the NIST database. For example,   Ir III, Pt III, and Au III as well as elements of $Z\geq 91$ are not contained. In such cases, one must rely on atomic structure codes to infer the energy levels. For instance, \cite{Gillanders2021MNRAS} studied the energy levels of Pt III and Au III by using \texttt{GRASP$^0$} and showed that there are no emission lines in $3\,{\rm \mu m}$ to $5\,{\rm \mu m}$ arising from the transitions between the lower-lying energy levels.
 Nevertheless, we should keep in mind that there can be additional lines.
 
Another caveat is that we do not include electric dipole transitions (E1).  This is a reasonable assumption for the typical ion, of which the energy scale of E1 transitions is usually larger than $1\,{\rm eV}$. However, some lanthanide ions, e.g, Ce II and Tb II, have E1 transition lines in mid-infrared. Including these lines may also affect our conclusion.
{Finally, we use collision strengths computed with \texttt{HULLAC}, which are not as accurate as those 
with the R-matrix method. For the transitions between the levels within the ground term of Se III and that between the ground and first excited level of W IV, we find that our collision strengths agree within a factor of $2$ with those  presented in \cite{Sterling2017ApJ} and \cite{Ballance2013JPhB}, where the R-matrix method is used. 
However, for Te III and  excitation to the higher levels of W IV, 
our collision strengths can be lower by a factor of $\sim 5$ \citep{Madonna2018ApJ,Ballance2013JPhB}. This difference may be due to 
the fact that we do not include resonant excitation. 
Therefore, the relative line intensities may  change if more accurate collision strengths are used.}


\section{Conclusion}
We study the kilonova nebular emission in the range of 3 -- 5 ${\rm \mu m}$, where the {\it Spitzer} Space Telescope observed the kilonova in GW170817 at 43 and 74 days \citep{Kasliwal2022MNRAS}. 
We  produce an M1 line list for $X^{+0}$--$X^{+3}$ from the energy levels  available in the NIST database \citep{NISTASD} by using the M1 selection rules in LS coupling under the single configuration approximation. This method enables to generate the synthetic spectra  in infrared with   accurate wavelengths.
In the range of 3 -- 5 ${\rm \mu m}$ our line list contains $\sim 100$ fine structure lines. 
With this list, the synthetic nebular spectra with two sets of the abundance patterns with or without the first r-process peak elements are computed for two different ionization states, $X^{+1}\&\,X^{+2}$ and $X^{+2}\&\,X^{+3}$.
Our findings and predictions are the followings:
\begin{itemize}
    \item The kilonova nebular spectrum in mid-infrared is sensitive to the abundance pattern whether the first r-process peak elements are included or not. Nevertheless, the synthetic spectra can match the observed $4.5\,{\rm \mu m}$ detection and $3.6\,{\rm \mu m}$ upper limit in both cases.
    \item If the first r-process peak elements are  abundant, ${\rm [Se\,III]\,}4.55\,{\rm \mu m}$ completely dominates the flux at the  $4.5\,{\rm \mu m}$ band because Se is the most abundant atomic species. In this case a strong emission feature due to ${\rm [Se\,III]\,}5.74\,{\rm \mu m}$ must exist, which may be accompanied by  [Ge II]\,5.66\,${\rm \mu m}$. 
    \item If the ejecta is dominated by elements beyond the first peak, W III, Os III, Rh II, Rh III, and Ce IV can produce an emission feature around $4.5\,{\rm \mu m}$. 
    Combining their emission lines can account for the observed flux  with the solar r-process abundances.
    \item If W III and Os III are the main sources of the observed flux  we expect that ${\rm [Os\, III]\,}2.60\,{\rm \mu m}$, ${\rm [Os\, II]\,}2.78\,{\rm \mu m}$, ${\rm [Os\,II]\,}6.01\,{\rm \mu m}$, and  ${\rm [W\,II]\,}6.05\,{\rm \mu m}$ may produce observable features.
    \item In the case without the first peak elements, some singly ionized ions, e.g,, Sb II and Ce II, have 
    strong lines around $3.6\,{\rm \mu m}$.
    The observed upper limit  implies that these atoms are likely in the higher ionization states. 
\end{itemize}

If a kilonova is successfully identified 
within $\sim 100\,{\rm Mpc}$ in the future,
the James Webb Space Telescope  will be able to provide the nebular spectra with a better quality across a wider wavelength range. With such data  more detailed analysis can be done. In order to extract more information about the elemental abundances and the ejecta properties, more atomic data, e.g., collision strengths and complete line lists \citep{Gaigalas2019ApJS,Gillanders2021MNRAS,Silva2022Atoms}, as well as  sophisticated modelings of the ionization states \citep{Hotokezaka2021MNRAS,Pognan2022MNRAS} are needed. 
\vspace{-0.5cm}
\section*{Acknowledgments}
We thank N. Domoto, K. Kawaguchi, and Y. Tarumi  for useful discussion and comments. We also thank M. Busquet for the generous support on the \texttt{HULLAC} code. This work was supported by Japan Society for the Promotion of Science (JSPS)  Early-Career Scientists Grant Number 20K14513, Grant-in-Aid for Scientific Research from JSPS (20H05639, 20H00158, 19H00694, 21H04997), and MEXT (17H06363).
\vspace{-0.5cm}
\section*{DATA AVAILABILITY}
The data presented this article will be shared on  request to the corresponding author.

\bibliographystyle{mnras}
\bibliography{ref.bib}

\begin{thebibliography}{}
\makeatletter
\relax
\def\mn@urlcharsother{\let\do\@makeother \do\$\do\&\do\#\do\^\do\_\do\%\do\~}
\def\mn@doi{\begingroup\mn@urlcharsother \@ifnextchar [ {\mn@doi@}
  {\mn@doi@[]}}
\def\mn@doi@[#1]#2{\def\@tempa{#1}\ifx\@tempa\@empty \href
  {http://dx.doi.org/#2} {doi:#2}\else \href {http://dx.doi.org/#2} {#1}\fi
  \endgroup}
\def\mn@eprint#1#2{\mn@eprint@#1:#2::\@nil}
\def\mn@eprint@arXiv#1{\href {http://arxiv.org/abs/#1} {{\tt arXiv:#1}}}
\def\mn@eprint@dblp#1{\href {http://dblp.uni-trier.de/rec/bibtex/#1.xml}
  {dblp:#1}}
\def\mn@eprint@#1:#2:#3:#4\@nil{\def\@tempa {#1}\def\@tempb {#2}\def\@tempc
  {#3}\ifx \@tempc \@empty \let \@tempc \@tempb \let \@tempb \@tempa \fi \ifx
  \@tempb \@empty \def\@tempb {arXiv}\fi \@ifundefined
  {mn@eprint@\@tempb}{\@tempb:\@tempc}{\expandafter \expandafter \csname
  mn@eprint@\@tempb\endcsname \expandafter{\@tempc}}}

\bibitem[\protect\citeauthoryear{{Abbott} et~al.,}{{Abbott}
  et~al.}{2017}]{Abbott2017ApJ}
{Abbott} B.~P.,  et~al., 2017, \mn@doi [\apjl] {10.3847/2041-8213/aa91c9},
  \href {https://ui.adsabs.harvard.edu/abs/2017ApJ...848L..12A} {848, L12}

\bibitem[\protect\citeauthoryear{{Azarov}, {Tchang-Brillet}, {Wyart}  \&
  {Meijer}}{{Azarov} et~al.}{2003}]{Azarov2003PhyS}
{Azarov} V.~I.,  {Tchang-Brillet} W. {\"U}.~L.,  {Wyart} J.~F.,   {Meijer}
  F.~G.,  2003, \mn@doi [\physscr] {10.1238/Physica.Regular.067a00190}, \href
  {https://ui.adsabs.harvard.edu/abs/2003PhyS...67..190A} {67, 190}

\bibitem[\protect\citeauthoryear{{Bahcall} \& {Wolf}}{{Bahcall} \&
  {Wolf}}{1968}]{Bahcall1968ApJ}
{Bahcall} J.~N.,  {Wolf} R.~A.,  1968, \mn@doi [\apj] {10.1086/149589}, \href
  {https://ui.adsabs.harvard.edu/abs/1968ApJ...152..701B} {152, 701}

\bibitem[\protect\citeauthoryear{{Ballance}, {Loch}, {Pindzola}  \&
  {Griffin}}{{Ballance} et~al.}{2013}]{Ballance2013JPhB}
{Ballance} C.~P.,  {Loch} S.~D.,  {Pindzola} M.~S.,   {Griffin} D.~C.,  2013,
  \mn@doi [Journal of Physics B Atomic Molecular Physics]
  {10.1088/0953-4075/46/5/055202}, \href
  {https://ui.adsabs.harvard.edu/abs/2013JPhB...46e5202B} {46, 055202}

\bibitem[\protect\citeauthoryear{{Bar-Shalom}, {Klapisch}  \&
  {Oreg}}{{Bar-Shalom} et~al.}{2001}]{Bar-Shalom2001JQSR}
{Bar-Shalom} A.,  {Klapisch} M.,   {Oreg} J.,  2001, \mn@doi [\jqsrt]
  {10.1016/S0022-4073(01)00066-8}, \href
  {https://ui.adsabs.harvard.edu/abs/2001JQSRT..71..169B} {71, 169}

\bibitem[\protect\citeauthoryear{{Barnes} \& {Kasen}}{{Barnes} \&
  {Kasen}}{2013}]{barnes2013ApJ}
{Barnes} J.,  {Kasen} D.,  2013, \mn@doi [\apj] {10.1088/0004-637X/775/1/18},
  \href {http://adsabs.harvard.edu/abs/2013ApJ...775...18B} {775, 18}

\bibitem[\protect\citeauthoryear{{Barnes}, {Kasen}, {Wu}  \&
  {Mart{\'\i}nez-Pinedo}}{{Barnes} et~al.}{2016}]{Barnes2016ApJ}
{Barnes} J.,  {Kasen} D.,  {Wu} M.-R.,   {Mart{\'\i}nez-Pinedo} G.,  2016,
  \mn@doi [\apj] {10.3847/0004-637X/829/2/110}, \href
  {https://ui.adsabs.harvard.edu/abs/2016ApJ...829..110B} {829, 110}

\bibitem[\protect\citeauthoryear{{Berrington}}{{Berrington}}{1995}]{Berrington1995}
{Berrington} K.~A.,  1995, \aaps, \href
  {https://ui.adsabs.harvard.edu/abs/1995A&AS..109..193B} {109, 193}

\bibitem[\protect\citeauthoryear{{Burbidge}, {Burbidge}, {Fowler}  \&
  {Hoyle}}{{Burbidge} et~al.}{1957}]{Burbidge1957RvMPB}
{Burbidge} E.~M.,  {Burbidge} G.~R.,  {Fowler} W.~A.,   {Hoyle} F.,  1957,
  \mn@doi [Reviews of Modern Physics] {10.1103/RevModPhys.29.547}, \href
  {https://ui.adsabs.harvard.edu/abs/1957RvMP...29..547B} {29, 547}

\bibitem[\protect\citeauthoryear{{Cameron}}{{Cameron}}{1957}]{Cameron1957PASP}
{Cameron} A.~G.~W.,  1957, \mn@doi [\pasp] {10.1086/127051}, \href
  {https://ui.adsabs.harvard.edu/abs/1957PASP...69..201C} {69, 201}

\bibitem[\protect\citeauthoryear{{Domoto}, {Tanaka}, {Wanajo}  \&
  {Kawaguchi}}{{Domoto} et~al.}{2021}]{Domoto2021ApJ}
{Domoto} N.,  {Tanaka} M.,  {Wanajo} S.,   {Kawaguchi} K.,  2021, \mn@doi
  [\apj] {10.3847/1538-4357/abf358}, \href
  {https://ui.adsabs.harvard.edu/abs/2021ApJ...913...26D} {913, 26}

\bibitem[\protect\citeauthoryear{{Fontes}, {Fryer}, {Hungerford}, {Wollaeger}
  \& {Korobkin}}{{Fontes} et~al.}{2020}]{Fontes2020MNRAS}
{Fontes} C.~J.,  {Fryer} C.~L.,  {Hungerford} A.~L.,  {Wollaeger} R.~T.,
  {Korobkin} O.,  2020, \mn@doi [\mnras] {10.1093/mnras/staa485}, \href
  {https://ui.adsabs.harvard.edu/abs/2020MNRAS.493.4143F} {493, 4143}

\bibitem[\protect\citeauthoryear{{Gaigalas}, {Kato}, {Rynkun},
  {Rad{\v{z}}i{\={u}}t{\.{e}}}  \& {Tanaka}}{{Gaigalas}
  et~al.}{2019}]{Gaigalas2019ApJS}
{Gaigalas} G.,  {Kato} D.,  {Rynkun} P.,  {Rad{\v{z}}i{\={u}}t{\.{e}}} L.,
  {Tanaka} M.,  2019, \mn@doi [\apjs] {10.3847/1538-4365/aaf9b8}, \href
  {https://ui.adsabs.harvard.edu/abs/2019ApJS..240...29G} {240, 29}

\bibitem[\protect\citeauthoryear{{Gall}, {Hjorth}, {Rosswog}, {Tanvir}  \&
  {Levan}}{{Gall} et~al.}{2017}]{Gall2017ApJ}
{Gall} C.,  {Hjorth} J.,  {Rosswog} S.,  {Tanvir} N.~R.,   {Levan} A.~J.,
  2017, \mn@doi [\apjl] {10.3847/2041-8213/aa93f9}, \href
  {https://ui.adsabs.harvard.edu/abs/2017ApJ...849L..19G} {849, L19}

\bibitem[\protect\citeauthoryear{{Gillanders}, {McCann}, {Sim}, {Smartt}  \&
  {Ballance}}{{Gillanders} et~al.}{2021}]{Gillanders2021MNRAS}
{Gillanders} J.~H.,  {McCann} M.,  {Sim} S.~A.,  {Smartt} S.~J.,   {Ballance}
  C.~P.,  2021, \mn@doi [\mnras] {10.1093/mnras/stab1861}, \href
  {https://ui.adsabs.harvard.edu/abs/2021MNRAS.506.3560G} {506, 3560}

\bibitem[\protect\citeauthoryear{{Gillanders}, {Smartt}, {Sim}, {Bauswein}  \&
  {Goriely}}{{Gillanders} et~al.}{2022}]{Gillanders2022}
{Gillanders} J.~H.,  {Smartt} S.~J.,  {Sim} S.~A.,  {Bauswein} A.,   {Goriely}
  S.,  2022, arXiv e-prints, \href
  {https://ui.adsabs.harvard.edu/abs/2022arXiv220201786G} {p. arXiv:2202.01786}

\bibitem[\protect\citeauthoryear{{Goriely}}{{Goriely}}{1999}]{Goriely1999A&A}
{Goriely} S.,  1999, \aap, \href
  {https://ui.adsabs.harvard.edu/abs/1999A&A...342..881G} {342, 881}

\bibitem[\protect\citeauthoryear{{Hotokezaka} \& {Nakar}}{{Hotokezaka} \&
  {Nakar}}{2020}]{Hotokezaka2020ApJ}
{Hotokezaka} K.,  {Nakar} E.,  2020, \mn@doi [\apj] {10.3847/1538-4357/ab6a98},
  \href {https://ui.adsabs.harvard.edu/abs/2020ApJ...891..152H} {891, 152}

\bibitem[\protect\citeauthoryear{{Hotokezaka}, {Beniamini}  \&
  {Piran}}{{Hotokezaka} et~al.}{2018}]{Hotokezaka2018IJMPD}
{Hotokezaka} K.,  {Beniamini} P.,   {Piran} T.,  2018, \mn@doi [International
  Journal of Modern Physics D] {10.1142/S0218271818420051}, \href
  {https://ui.adsabs.harvard.edu/abs/2018IJMPD..2742005H} {27, 1842005}

\bibitem[\protect\citeauthoryear{{Hotokezaka}, {Tanaka}, {Kato}  \&
  {Gaigalas}}{{Hotokezaka} et~al.}{2021}]{Hotokezaka2021MNRAS}
{Hotokezaka} K.,  {Tanaka} M.,  {Kato} D.,   {Gaigalas} G.,  2021, \mn@doi
  [\mnras] {10.1093/mnras/stab1975}, \href
  {https://ui.adsabs.harvard.edu/abs/2021MNRAS.506.5863H} {506, 5863}

\bibitem[\protect\citeauthoryear{{Kasen} \& {Barnes}}{{Kasen} \&
  {Barnes}}{2019}]{Kasen2019ApJ}
{Kasen} D.,  {Barnes} J.,  2019, \mn@doi [\apj] {10.3847/1538-4357/ab06c2},
  \href {https://ui.adsabs.harvard.edu/abs/2019ApJ...876..128K} {876, 128}

\bibitem[\protect\citeauthoryear{{Kasen}, {Metzger}, {Barnes}, {Quataert}  \&
  {Ramirez-Ruiz}}{{Kasen} et~al.}{2017}]{Kasen2017}
{Kasen} D.,  {Metzger} B.,  {Barnes} J.,  {Quataert} E.,   {Ramirez-Ruiz} E.,
  2017, \mn@doi [\nat] {10.1038/nature24453}, \href
  {http://adsabs.harvard.edu/abs/2017Natur.551...80K} {551, 80}

\bibitem[\protect\citeauthoryear{{Kasliwal} et~al.,}{{Kasliwal}
  et~al.}{2022}]{Kasliwal2022MNRAS}
{Kasliwal} M.~M.,  et~al., 2022, \mn@doi [\mnras] {10.1093/mnrasl/slz007},
  \href {https://ui.adsabs.harvard.edu/abs/2022MNRAS.510L...7K} {510, L7}

\bibitem[\protect\citeauthoryear{{Kawaguchi}, {Shibata}  \&
  {Tanaka}}{{Kawaguchi} et~al.}{2018}]{kawaguchi2018}
{Kawaguchi} K.,  {Shibata} M.,   {Tanaka} M.,  2018, \mn@doi [\apjl]
  {10.3847/2041-8213/aade02}, \href
  {http://adsabs.harvard.edu/abs/2018ApJ...865L..21K} {865, L21}

\bibitem[\protect\citeauthoryear{Kramida, {Yu.~Ralchenko}, Reader  \& {and NIST
  ASD Team}}{Kramida et~al.}{2021}]{NISTASD}
Kramida A.,  {Yu.~Ralchenko} Reader J.,   {and NIST ASD Team} 2021, {NIST
  Atomic Spectra Database (ver. 5.9), [Online]. Available:
  {\tt{https://physics.nist.gov/asd}} [2017, April 9]. National Institute of
  Standards and Technology, Gaithersburg, MD.}

\bibitem[\protect\citeauthoryear{{Lattimer} \& {Schramm}}{{Lattimer} \&
  {Schramm}}{1974}]{Lattimer1974ApJ}
{Lattimer} J.~M.,  {Schramm} D.~N.,  1974, \mn@doi [\apjl] {10.1086/181612},
  \href {https://ui.adsabs.harvard.edu/abs/1974ApJ...192L.145L} {192, L145}

\bibitem[\protect\citeauthoryear{{Madonna} et~al.,}{{Madonna}
  et~al.}{2018}]{Madonna2018ApJ}
{Madonna} S.,  et~al., 2018, \mn@doi [\apjl] {10.3847/2041-8213/aaccef}, \href
  {https://ui.adsabs.harvard.edu/abs/2018ApJ...861L...8M} {861, L8}

\bibitem[\protect\citeauthoryear{{Malcheva} et~al.,}{{Malcheva}
  et~al.}{2009}]{Malcheva2009MNRAS}
{Malcheva} G.,  et~al., 2009, \mn@doi [\mnras]
  {10.1111/j.1365-2966.2009.14894.x}, \href
  {https://ui.adsabs.harvard.edu/abs/2009MNRAS.396.2289M} {396, 2289}

\bibitem[\protect\citeauthoryear{{Margutti} \& {Chornock}}{{Margutti} \&
  {Chornock}}{2021}]{Margutti2021ARA&A}
{Margutti} R.,  {Chornock} R.,  2021, \mn@doi [\araa]
  {10.1146/annurev-astro-112420-030742}, \href
  {https://ui.adsabs.harvard.edu/abs/2021ARA&A..59..155M} {59}

\bibitem[\protect\citeauthoryear{{Metzger}}{{Metzger}}{2017}]{Metzger2017LRR}
{Metzger} B.~D.,  2017, \mn@doi [Living Reviews in Relativity]
  {10.1007/s41114-017-0006-z}, \href
  {https://ui.adsabs.harvard.edu/abs/2017LRR....20....3M} {20, 3}

\bibitem[\protect\citeauthoryear{{Nakar}}{{Nakar}}{2020}]{Nakar2020PhR}
{Nakar} E.,  2020, \mn@doi [\physrep] {10.1016/j.physrep.2020.08.008}, \href
  {https://ui.adsabs.harvard.edu/abs/2020PhR...886....1N} {886, 1}

\bibitem[\protect\citeauthoryear{{Pasternack}}{{Pasternack}}{1940}]{Pasternack1940ApJ}
{Pasternack} S.,  1940, \mn@doi [\apj] {10.1086/144208}, \href
  {https://ui.adsabs.harvard.edu/abs/1940ApJ....92..129P} {92, 129}

\bibitem[\protect\citeauthoryear{{Perego} et~al.,}{{Perego}
  et~al.}{2022}]{Perego2022ApJ}
{Perego} A.,  et~al., 2022, \mn@doi [\apj] {10.3847/1538-4357/ac3751}, \href
  {https://ui.adsabs.harvard.edu/abs/2022ApJ...925...22P} {925, 22}

\bibitem[\protect\citeauthoryear{{Pognan}, {Jerkstrand}  \& {Grumer}}{{Pognan}
  et~al.}{2022a}]{Pognan2022}
{Pognan} Q.,  {Jerkstrand} A.,   {Grumer} J.,  2022a, arXiv e-prints, \href
  {https://ui.adsabs.harvard.edu/abs/2022arXiv220209245P} {p. arXiv:2202.09245}

\bibitem[\protect\citeauthoryear{{Pognan}, {Jerkstrand}  \& {Grumer}}{{Pognan}
  et~al.}{2022b}]{Pognan2022MNRAS}
{Pognan} Q.,  {Jerkstrand} A.,   {Grumer} J.,  2022b, \mn@doi [\mnras]
  {10.1093/mnras/stab3674}, \href
  {https://ui.adsabs.harvard.edu/abs/2022MNRAS.510.3806P} {510, 3806}

\bibitem[\protect\citeauthoryear{{Preval}, {Badnell}  \& {O'Mullane}}{{Preval}
  et~al.}{2019}]{Preval2019JPhB}
{Preval} S.~P.,  {Badnell} N.~R.,   {O'Mullane} M.~G.,  2019, \mn@doi [Journal
  of Physics B Atomic Molecular Physics] {10.1088/1361-6455/aaf3f4}, \href
  {https://ui.adsabs.harvard.edu/abs/2019JPhB...52b5201P} {52, 025201}

\bibitem[\protect\citeauthoryear{{Rosswog}, {Sollerman}, {Feindt}, {Goobar},
  {Korobkin}, {Wollaeger}, {Fremling}  \& {Kasliwal}}{{Rosswog}
  et~al.}{2018}]{Rosswog2018A&A}
{Rosswog} S.,  {Sollerman} J.,  {Feindt} U.,  {Goobar} A.,  {Korobkin} O.,
  {Wollaeger} R.,  {Fremling} C.,   {Kasliwal} M.~M.,  2018, \mn@doi [\aap]
  {10.1051/0004-6361/201732117}, \href
  {https://ui.adsabs.harvard.edu/abs/2018A&A...615A.132R} {615, A132}

\bibitem[\protect\citeauthoryear{{Shortley}}{{Shortley}}{1940}]{Shortly1940}
{Shortley} G.~H.,  1940, \mn@doi [Physical Review] {10.1103/PhysRev.57.225},
  \href {https://ui.adsabs.harvard.edu/abs/1940PhRv...57..225S} {57, 225}

\bibitem[\protect\citeauthoryear{{Silva}, {Sampaio}, {Amaro}, {Fl{\"o}rs},
  {Mart{\'\i}nez-Pinedo}  \& {Marques}}{{Silva} et~al.}{2022}]{Silva2022Atoms}
{Silva} R.~F.,  {Sampaio} J.~M.,  {Amaro} P.,  {Fl{\"o}rs} A.,
  {Mart{\'\i}nez-Pinedo} G.,   {Marques} J.~P.,  2022, \mn@doi [Atoms]
  {10.3390/atoms10010018}, \href
  {https://ui.adsabs.harvard.edu/abs/2022Atoms..10...18S} {10, 18}

\bibitem[\protect\citeauthoryear{{Sterling} \& {Witthoeft}}{{Sterling} \&
  {Witthoeft}}{2011}]{Sterling2011A&A}
{Sterling} N.~C.,  {Witthoeft} M.~C.,  2011, \mn@doi [\aap]
  {10.1051/0004-6361/201116718}, \href
  {https://ui.adsabs.harvard.edu/abs/2011A&A...529A.147S} {529, A147}

\bibitem[\protect\citeauthoryear{{Sterling}, {Madonna}, {Butler},
  {Garc{\'\i}a-Rojas}, {Mashburn}, {Morisset}, {Luridiana}  \&
  {Roederer}}{{Sterling} et~al.}{2017}]{Sterling2017ApJ}
{Sterling} N.~C.,  {Madonna} S.,  {Butler} K.,  {Garc{\'\i}a-Rojas} J.,
  {Mashburn} A.~L.,  {Morisset} C.,  {Luridiana} V.,   {Roederer} I.~U.,  2017,
  \mn@doi [\apj] {10.3847/1538-4357/aa6c28}, \href
  {https://ui.adsabs.harvard.edu/abs/2017ApJ...840...80S} {840, 80}

\bibitem[\protect\citeauthoryear{{Takami}, {Nozawa}  \& {Ioka}}{{Takami}
  et~al.}{2014}]{Takami2014ApJ}
{Takami} H.,  {Nozawa} T.,   {Ioka} K.,  2014, \mn@doi [\apjl]
  {10.1088/2041-8205/789/1/L6}, \href
  {https://ui.adsabs.harvard.edu/abs/2014ApJ...789L...6T} {789, L6}

\bibitem[\protect\citeauthoryear{{Tanaka} \& {Hotokezaka}}{{Tanaka} \&
  {Hotokezaka}}{2013}]{tanaka2013}
{Tanaka} M.,  {Hotokezaka} K.,  2013, \mn@doi [\apj]
  {10.1088/0004-637X/775/2/113}, \href
  {https://ui.adsabs.harvard.edu/abs/2013ApJ...775..113T} {775, 113}

\bibitem[\protect\citeauthoryear{{Tanaka} et~al.,}{{Tanaka}
  et~al.}{2017}]{Tanaka2017PASJ}
{Tanaka} M.,  et~al., 2017, \mn@doi [\pasj] {10.1093/pasj/psx121}, \href
  {https://ui.adsabs.harvard.edu/abs/2017PASJ...69..102T} {69, 102}

\bibitem[\protect\citeauthoryear{{Tanaka}, {Kato}, {Gaigalas}  \&
  {Kawaguchi}}{{Tanaka} et~al.}{2020}]{Tanaka2020MNRAS}
{Tanaka} M.,  {Kato} D.,  {Gaigalas} G.,   {Kawaguchi} K.,  2020, \mn@doi
  [\mnras] {10.1093/mnras/staa1576}, \href
  {https://ui.adsabs.harvard.edu/abs/2020MNRAS.496.1369T} {496, 1369}

\bibitem[\protect\citeauthoryear{{Villar} et~al.,}{{Villar}
  et~al.}{2018}]{Villar2018ApJ}
{Villar} V.~A.,  et~al., 2018, \mn@doi [\apjl] {10.3847/2041-8213/aad281},
  \href {https://ui.adsabs.harvard.edu/abs/2018ApJ...862L..11V} {862, L11}

\bibitem[\protect\citeauthoryear{{Watson} et~al.,}{{Watson}
  et~al.}{2019}]{Watson2019Natur}
{Watson} D.,  et~al., 2019, \mn@doi [\nat] {10.1038/s41586-019-1676-3}, \href
  {https://ui.adsabs.harvard.edu/abs/2019Natur.574..497W} {574, 497}

\bibitem[\protect\citeauthoryear{{Waxman}, {Ofek}  \& {Kushnir}}{{Waxman}
  et~al.}{2019}]{Waxman2019ApJ}
{Waxman} E.,  {Ofek} E.~O.,   {Kushnir} D.,  2019, \mn@doi [\apj]
  {10.3847/1538-4357/ab1f71}, \href
  {https://ui.adsabs.harvard.edu/abs/2019ApJ...878...93W} {878, 93}

\bibitem[\protect\citeauthoryear{{Wollaeger} et~al.,}{{Wollaeger}
  et~al.}{2021}]{Wollaeger2021ApJ}
{Wollaeger} R.~T.,  et~al., 2021, \mn@doi [\apj] {10.3847/1538-4357/ac0d03},
  \href {https://ui.adsabs.harvard.edu/abs/2021ApJ...918...10W} {918, 10}

\makeatother
\end{thebibliography}
\bsp	
\label{lastpage}
\end{document}